\documentclass[showpacs,preprintnumbers,amsmath,amssymb]{revtex4}
\usepackage{amsmath,amsfonts,latexsym,amssymb,graphicx,graphics,epsfig,subfigure,color,makeidx}
\usepackage{xcolor,diagbox}
\usepackage{multirow}
\usepackage[colorlinks,linkcolor=blue,anchorcolor=blue,citecolor=green,urlcolor=blue]{hyperref}

\usepackage[latin1]{inputenc}

\newcommand {\nn}    {\nonumber}
\newcommand {\fc}    {\frac}
\newcommand {\be}    {\begin{equation}}
\newcommand {\ee}    {\end{equation}}
\newcommand {\beq}   {\begin{eqnarray}}
\newcommand {\eeq}   {\end{eqnarray}}

\newcommand {\lt}    {\left}
\newcommand {\rt}    {\right}

\begin{document}

\title{Quasinormal modes and greybody factor of a Lorentz-violating black hole}

\author{Wen-Di Guo$^{ab}$\footnote{guowd@lzu.edu.cn}}
\author{Qin Tan$^{c}$\footnote{tanqin@hunnu.edu.cn}}
\author{Yu-Xiao Liu$^{ab}$\footnote{liuyx@lzu.edu.cn, corresponding author}}

\affiliation{$^{a}$Lanzhou Center for Theoretical Physics, Key Laboratory of Theoretical Physics of Gansu Province,
School of Physical Science and Technology, Lanzhou University, Lanzhou 730000, China\\
             $^{b}$Institute of Theoretical Physics $\&$ Research Center of Gravitation, Lanzhou University, Lanzhou 730000, China\\
             $^{c}$Department of Physics, Key Laboratory of Low Dimensional Quantum Structures and Quantum Control of Ministry of Education, Synergetic Innovation Center for Quantum Effects and Applications, Hunan Normal University, Changsha, 410081, Hunan, China}

\begin{abstract}
Recently, a static spherically symmetric black hole solution was found in gravity nonminimally coupled a background Kalb-Ramond field. The Lorentz symmetry is spontaneously broken when the Kalb-Ramond field has a nonvanishing vacuum expectation value. In this work, we focus on the quasinormal modes and greybody factor of this black hole. The master equations for the perturbed scalar field, electromagnetic field, and gravitational field can be written into a Schr\"{o}dinger equation. We use three methods to solve the quasinormal frequencies in the frequency domain. The results agree well with each other. The time evolution of a Gaussian wave packet is studied. The quasinormal frequencies fitted from the time evolution data agree well with that of frequency domain. The greybody factor is calculated by Wentzel-Kramers-Brillouin (WKB) method. The effect of the Lorentz-violating parameter on the quasinormal modes and greybody factor are also studied.
\end{abstract}



\maketitle

\section{Introduction}

Lorentz symmetry is a fundamental concept in modern physics. It states that all physical laws are same in all inertial frames. Although Lorentz symmetry has been thought to be very important and has been tested by a lot of experiments and observations, it also can be violated in some gravity theories, for example, Einstein-{\ae}ther theory~\cite{Jacobson:2000xp}, Horava-Lifshitz gravity~\cite{Horava:2009uw}, $f(T)$ gravity~\cite{Bengochea:2008gz}, or in some high energy scale, such as string theory~\cite{Kostelecky:1988zi} and loop quantum gravity~\cite{Alfaro:2001rb}. One of the frameworks for the Lorentz violating theories is the Standard-Model-Extension~\cite{Kostelecky:2003fs}. Different from the explicitly broken Lorentz symmetry theories, in this framework, the Lagrangian density is Lorentz invariant, but the ground state of the system will not. This can be achieved by some background field acquiring a nonzero vacuum expectation value.

One of the famous theories is the bumblebee gravity~\cite{Kostelecky:1989jw,Kostelecky:1989jp,Bailey:2006fd,Bluhm:2008yt}. A vector field named as the bumblebee field with a nonzero vacuum expectation value selects a specific direction, this will break the local particle Lorentz symmetry. The bumblebee model has been studied widely. A static spherical symmetric black hole solution was found by Casana et al. in Ref.~\cite{Casana:2017jkc}. Then it was generalized to (anti) de Sitter cases~\cite{Maluf:2020kgf}. Two families  of static spherical black hole solutions were obtained by investigating the background bumblebee field with a non-vanishing temporal component or radial component~\cite{Xu:2022frb}. The rotating bumblebee black holes were studied in Refs.~\cite{Ding:2019mal,Ding:2020kfr}. A static spherical black hole solution with a global monopole was proposed in Ref.~\cite{Gullu:2020qzu}. Other black hole solutions were also studied~\cite{Ding:2021iwv,Jha:2020pvk,Ding:2022qcy}. The thermodynamic properties and observation effects of the bumblebee black holes were investigated in Refs.~\cite{Mai:2023ggs,Wang:2021irh,Chen:2023cjd,Zhang:2023wwk,Lin:2023foj,Wang:2021gtd,Duan:2023gng}.

Instead the bumblebee field, the local particle Lorentz symmetry can also be broken by a rank-two antisymmetric tensor field, named as the Kalb-Ramond (KR) field ~\cite{Altschul:2009ae}. The KR field can emerge from string theory~\cite{Kalb:1974yc}. When the KR field has a nonzero vacuum expectation value and couples to gravity, the Lorentz symmetry can be broken spontaneously. Under the nonzero vacuum expectation value, a static and spherically symmetric solution was constructed in Ref.~\cite{Lessa:2019bgi}. Recently, a new class of solutions with and without the cosmological constant were proposed by Yang et al. in Ref.~\cite{Yang:2023wtu}. The shadow and the quasinormal modes (QNMs) of this black hole were studied~\cite{Filho:2023ycx}.

The detection of the gravitational waves by Laser Interferometer Gravitational-Wave Observatory (LIGO) and Virgo~\cite{LIGOScientific:2016aoc} and the first picture of the black hole by Event Horizon Telescope (EHT)~\cite{EventHorizonTelescope:2019dse} label that we have entered the multi-messenger astronomy. The study on the gravitational waves and black hole shadows have been attracted a lot of attentions~\cite{Zeng:2021mok,Chen:2022qrw,Liu:2022csl,Chen:2022scf,Ali:2023moi,Li:2023plm,Zhao:2021zlr,Yi:2023mbm}.  For the gravitational waves of a binary black hole merger system, there are three parts: inspiral, merger, and ringdown. It is believed that the gravitational waves in the ringdown stage are dominated by the QNMs~\cite{Berti:2007dg}. The QNMs are characteristic modes of dissipative systems. Different from the normal modes, the QNMs do not form a complete set~\cite{Nollert:1998ys}. The real parts of the quasinormal frequencies (QNFs) are the oscillation frequencies of the perturbation, and the imaginary parts are related to the decay time. The QNMs of the gravitational field can only depend on the parameter of the black hole, so it can be used to infer the mass and angular momentum of a black hole, and further to test the no-hair theorem~\cite{Echeverria:1989hg,Berti:2005ys,Berti:2007zu,Isi:2019aib}. In addition, the ultracompact objects without event horizon could have echo signals, which could be used to test the existence of event horizons~\cite{Cardoso:2016rao,Cardoso:2019rvt,Cardoso:2017cqb}. Recently, it was found that the spectrum of the QNMs is unstable under the perturbation of the effective potential~\cite{Jaramillo:2020tuu,Cheung:2021bol}. The time domain observations and the corresponding Regge Poles were studied in Refs~\cite{Berti:2022xfj,Torres:2023nqg}. The QNMs in modified gravity theories were widely studied in Refs.~\cite{Guo:2021enm,Cardoso:2020nst,Guo:2022rms,Yang:2023gas,Guo:2023vmc,Li:2021ngc,Uniyal:2022xnq,Campos:2023zmg}. Besides, the QNMs in brane world models were also studied in Refs.~\cite{Seahra:2005wk,Tan:2022vfe}.

Another important concept for a black hole perturbation system is the greybody factor, which describes the transmission probability of an outgoing wave reach to infinity or an incoming wave to be absorbed by the black hole~\cite{Konoplya:2019ppy,Cardoso:2005vb,Dey:2018cws}. The greybody factor can describe the information of the near horizon regions of a black hole~\cite{Kanti:2002nr}. It can be used to evaluate the Hawking radiation energy emission~\cite{Hawking:1975vcx}. Recently, it was pointed out that the ringdown signal after an extreme mass ratio merger can be modelled by the greybody factor~\cite{Oshita:2023cjz}.

In this paper we want to study the QNMs and greybody factor of the scalar field, electromagnetic field and gravitational field in the background of the Lorentz-violating black hole in gravity nonminimally coupled a background Kalb-Ramond field. This paper is organized as follows. In Sec.~\ref{KR_bh}, we review the black hole solution briefly and give the master equations for the perturbed fields. In Sec.~\ref{QNM}, we solve the QNFs with three different methods, and study the effect of the Lorentz-violating parameter on the QNMs. In Sec.~\ref{gbfactor}, we calculate the greybody factor with the Wentzel-Kramers-Brillouin (WKB) method. The conclusions are given in Sec.~\ref{conclusion}.

\section{Background and perturbation equations}\label{KR_bh}
In this section we review the Lorentz-violating black hole briefly and derive the perturbed master equations for the scalar field, electromagnetic field, and gravitational field. The starting point is the following Einstein-Hilbert action coupled with a self-interacting KR field non-minimally~\cite{Altschul:2009ae,Lessa:2019bgi}
\be
S=\fc{1}{2\kappa}\int d^4x\sqrt{-g}\lt[R-\fc{1}{6}H^{\mu\nu\rho}H_{\mu\nu\rho}-V(B^{\mu\nu}B_{\mu\nu}\pm b^2)+\xi B^{\rho\mu}B^{\nu}_{\mu}R_{\rho\nu} \rt]+\int d^4x \sqrt{-g}\mathcal{L}_M,
\ee
where $\kappa$ is related to the Newtonian constant $G$ through $\kappa=8\pi G$, $\xi$ is the coupling constant between the KR field and the gravity. The field strength of the KR field is defined as $H_{\mu\nu\rho}\equiv\partial_{[\mu}B_{\nu\rho]}$. The key point is the nonvanishing vacuum expectation value (VEV) for the KR field, $\langle B_{\mu\nu}\rangle=b_{\mu\nu}$, and $b^{\mu\nu}b_{\mu\nu}=\mp b^2$. The sign $\pm$ in the potential $V$ is to ensure that the constant $b^2$ is a positive number~\cite{Altschul:2009ae,Kalb:1974yc,Bluhm:2007bd}. Due to the vacuum condensation, the gauge invariance of the KR field is spontaneously broken. Because the KR field is nonminimally coupled to the gravity, the spontaneous symmetry breaking could lead to the violation of local Lorentz symmetry. Similar to the Maxwell field, the KR field could be decomposed to a pseudo-electric and a pseudo-magnetic field. Assuming that only the pseudo-electric nonvanishing, the KR field strength will vanish, i.e., $H_{\lambda\mu\nu}=0$. Under the VEV configuration, the field equation could be written as
\beq
R_{\mu\nu}&=&V'(2b_{\mu\alpha}b_{\nu}^{\alpha}+b^2g_{\mu\nu})+\xi\Big[g_{\mu\nu}b^{\alpha\gamma}b^{\beta}_{\gamma}R_{\alpha\beta} -b^{\alpha}_{\mu}b^{\beta}_{\nu}R_{\alpha\beta}-b^{\alpha\beta}b_{\mu\beta}R_{\nu\alpha}-b^{\alpha\beta}b_{\nu\beta}R_{\mu\alpha}\nn\\
&+&\fc{1}{2}\nabla_{\alpha}\nabla_{\mu}(b^{\alpha\beta}b_{\nu\beta})+\fc{1}{2}\nabla_{\alpha}\nabla_{\nu}(b^{\alpha\beta}b_{\mu\beta}) -\fc{1}{2}\nabla^{\alpha}\nabla_{\alpha}(b_{\mu}^{\gamma}b_{\nu\gamma})\Big],
\eeq
where the prime denotes the derivative with respect to the corresponding argument. Under the assumption that the VEV locates at the minimum of the potential, that is $V'=0$, a Schwarzschild-like black hole solution in the theory is obtained~\cite{Yang:2023wtu}
\be
ds^2=-f(r)dt^2+\fc{dr^2}{f(r)}+r^2d\theta^2+r^2\sin^2\theta d\phi^2,\label{metricKR}
\ee
where $f(r)=\fc{1}{1-\gamma}-\fc{2M}{r}$ and $M$ is the Komar mass. The parameter $\gamma$ is dimensionless which is defined as $\gamma\equiv\xi b^2/2$. The Lorentz symmetry violation effect caused by the nonvanishing VEV of the KR field is characterized by the parameter $\gamma$. The event horizon locates at $r_h=2(1-\gamma)M$, which is different from the Schwarzschild black hole. Note that, in the limit $r\rightarrow\infty$, the ($\theta,\theta$) and ($\phi,\phi$) components of the Ricci tensor are $\fc{\gamma}{\gamma-1}$ and $\fc{\gamma\sin^2\theta}{\gamma-1}$, respectively. So the background spacetime is not asymptotically Minkowski for $\gamma\neq0$. Besides, the Kretschmann scalar $R^{\alpha\beta\mu\nu}R_{\alpha\beta\mu\nu}$ is~\cite{Yang:2023wtu}
\be
R^{\alpha\beta\gamma\delta}R_{\alpha\beta\gamma\delta}=\fc{48M^2}{r^6}-\fc{16\gamma M}{(1-\gamma)r^5}+\fc{4\gamma^2}{(1-\gamma)^2r^4}.
\ee
That means the Lorentz-violating effect cannot be removed by a coordinate transformation. The solutions with the presence of cosmological constant were also obtained in Ref.~\cite{Yang:2023wtu}.

With this background spacetime, we would like to study the scattering wave of a perturbed field $\Psi^s$, where $s$ is the spin of the perturbed field, i.e., $s=0,\pm1,\pm2$ correspond to the scalar field, electromagnetic field, and gravitational field, respectively. While the theoretical framework we employ permits departures from local Lorentz invariance, it is important to note that the specific perturbations examined within our paper indeed preserve Lorentz invariance. This stems from the fact that all the perturbed fields we consider are test fields, which do not interact with the KR field. Consequently, these test fields maintain local Lorentz invariance, and their propagation speed remains unaltered, consistent with the speed of light. The variable decomposition can be done with the help of the spherical harmonics, vectorial harmonics, and tensorial harmonics~\cite{Edmonds,Regge}. The perturbed field equation for the radial part in the tortoise coordinate can be written as
\be
\fc{d^2}{dr_*^2}\Psi^s+(\omega^2-V_{s})\Psi^s=0,\label{master equation}
\ee
where the tortoise coordinate is defined as $dr_*=\fc{dr}{f(r)}$. The effective potentials for the scalar field and electromagnetic field are
\begin{equation}\label{effctive potential}
V_{s}=
f(r)\lt(\fc{l(l+1)}{r^2}-s(s-1)\fc{\gamma^2}{r^2(1-\gamma)^2}+(1-s^2)\fc{2M}{r^3}\rt).
\end{equation}
For the gravitational perturbations, the effective potentials for the odd parity and even parity are
\begin{equation}\label{effctive potential gravodd}
V_s^{\text{~o}}=
f(r)\lt(\fc{(\mu+2)\eta^2-2(1-\eta)^2}{r^2\eta(2\eta-1)}-\fc{6M}{r^3}\rt),
\end{equation}
and
\begin{equation}\label{effctive potential graveven}
V_s^{\text{~e}}=
f(r)\lt(\fc{72\eta^3 M^3+36\eta^2 M^2 \mu r+6\eta M \mu^2 r^2+(\mu-2)\mu^2 r^3}{r^3\eta(6\eta M+\mu r)^2}\rt),
\end{equation}
respectively. Here $s=\pm2$ for the gravitational perturbations. We have denoted $\mu\equiv(l+2)(l-1)$, and $\eta\equiv1-\gamma$.

The derivation of Eq.~\eqref{master equation} is similar to the case of the Schwarzschild black hole~\cite{Regge}. Note that, for the scalar field ($s=0$) and electromagnetic field ($s=\pm1$), the form of the effective potential is same as that of the Schwarzschild black hole. The reason is that the field equations of the scalar field and electromagnetic field are not modified by the KR field.

\begin{figure*}[htb]
\begin{center}
\subfigure[$ $]  {\label{Vs}
\includegraphics[width=5.7cm]{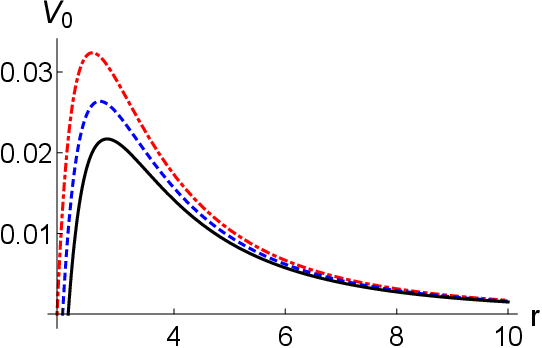}}
\subfigure[$ $]  {\label{VE}
\includegraphics[width=5.7cm]{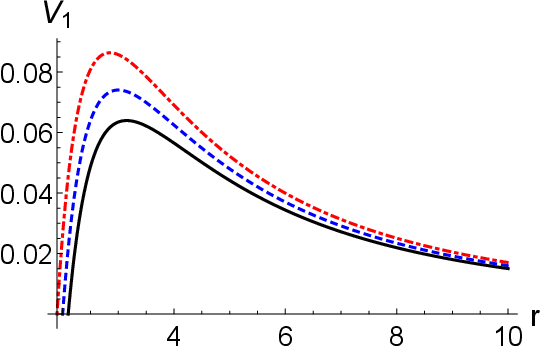}}
\subfigure[$ $]  {\label{Vg}
\includegraphics[width=5.7cm]{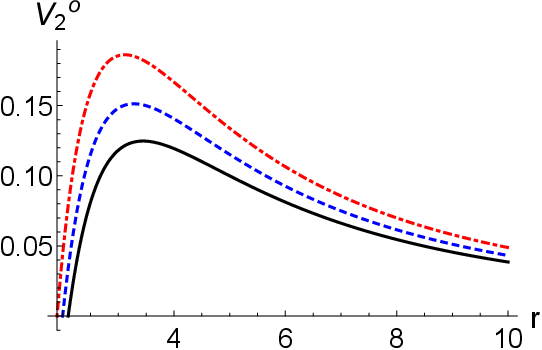}}
\end{center}
\caption{The effective potentials for different perturbed fields. The mass $M$ is set to $M=1$, and the Lorentz violating parameter $\gamma$ is set to $\gamma=-0.05$ (the black lines), $\gamma=0$ (the blue dots), and $\gamma=0.05$ (the red dashed dots).  (a) The effective potential for the scalar field and $l=0$. (b) The effective potential for the Maxwell field and $l=1$. (c)  The effective potential for the odd parity gravitational field and $l=2$.}
\label{plotpotential}
\end{figure*}

The plots for the effective potentials for the three perturbed fields are shown in Fig.~\ref{plotpotential}. The difference between these effective potentials for the odd parity and even parity gravitational perturbations is very small, so we only show that for the odd parity. From this figure we can see that, the effect of the parameter $\gamma$ on the effective potentials for the three perturbed fields are the same, that is the heights of the potentials increase with $\gamma$.

\section{Quasinormal mode frequencies}\label{QNM}

In this section, we will solve the master equation~\eqref{master equation} to obtain the QNFs with three different methods, the continued fraction method~\cite{Leaver1985}, the direct integration method~\cite{Pani:2013pma}, and the WKB method~\cite{Konoplya:2003ii}. Hereafter, we will set the mass $M=1$ in this paper. The boundary condition for this problem is
\beq
\Psi^s&\sim& e^{-i\omega r_*},~~~~~~r_*\rightarrow -\infty,\nn\\
\Psi^s&\sim& e^{i\omega r_*},~~~~~~~~r_*\rightarrow +\infty,\label{boundary}
\eeq
which is only an ingoing wave at the horizon and only an outgoing wave at infinity.

\emph{Continued fraction method} The continued fraction method proposed by Leaver~\cite{Leaver1985} is the most exact method to solve QNFs of a black hole. The wave function satisfying the boundary~\eqref{boundary} could be written as
\be
\Psi^s=(r-2\eta)^{-2i\omega \eta^2}r^{4i\omega \eta^2}e^{i\eta\omega(r-2\eta)}\sum_{n=0}a_n\fc{(r-2\eta)^n}{r^n},
\ee
where $a_n$ is the expansion coefficient. Substituting this into the master equation~\eqref{master equation}, a three-term recurrence relation can be obtained for the scalar field, electric field, and odd parity gravitational field:
\beq
\alpha_0a_1+\beta_0a_0&=&0, \label{alpha1}\\
\alpha_n a_{n+1}+\beta_n a_n+\delta_n a_{n-1}&=&0,~~~~~~n=1,2\ldots\label{alpha2}
\eeq
where the recurrence coefficients $\alpha_n$, $\beta_n$, and $\gamma_n$ are functions of $n$ and the parameters of the black hole:
\beq
\alpha_n=(n+1)(n+1-4i\eta^2\omega),
\eeq
\begin{displaymath}
\beta_n= \left\{ \begin{array}{ll}
-1-\eta l(l+1)-2n-2n^2+s^2+32\eta^4\omega^2+8i \eta^2(\omega+2n\omega) & \textrm{for the scalar field and electric field,}\\
-1 + \fc{2}{\eta} - \eta (-2 + l + l^2) - 2 n - 2 n^2 + 32 a^4 \omega^2 +
 8 i \eta^2 (\omega + 2 n \omega) & \textrm{for the odd parity gravitational field.}
\end{array} \right.
\end{displaymath}
\begin{displaymath}
\delta_n= \left\{ \begin{array}{ll}
(n-1)^2-s^2+(n-1)(2-8i\eta^2\omega)-(i+4a^2\omega)^2 & \textrm{for the scalar field and electric field,}\\
-4+n^2-8i\eta^2n\omega-16\eta^4\omega^2 & \textrm{for the odd parity gravitational field.}
\end{array} \right.
\end{displaymath}
For the even parity gravitational field, a five-term recurrence relation is obtained as ,
\beq
\alpha^e_0a_1+\beta^e_0a_0&=&0, \label{alphae1} \\
\alpha^e_1 a_{2}+\beta^e_1 a_1+\delta^e_1 a_{0}&=&0,\label{alphae2}\\
\alpha^e_2 a_{3}+\beta^e_2 a_2+\delta^e_2 a_{1}+\sigma^e_2 a_{0}&=&0,\label{alphae3}\\
\alpha^e_n a_{n+1}+\beta^e_n a_n+\delta^e_n a_{n-1}+\sigma^e_n a_{n-2}+\rho^e_n a_{n-3}&=&0,~~~~~~n=3,4\ldots\label{alphae4}
\eeq
with
\beq
\alpha^e_{n}&=&(\mu +1)^2 (n+1) \left(-4 i \eta ^2 \omega +n+1\right),\nn\\
\beta^e_{n}&=&(-\mu -1) \Big[-32 \eta ^4 \omega ^2-8 i \eta ^2 \omega -2 \mu  \left(16 \eta ^4 \omega ^2+4 i \eta ^2 \omega +1\right)+\mu ^2+2 (\mu +4) n^2\nn\\
&+&2 n \left(-8 i \eta ^2 \mu  \omega -20 i \eta ^2 \omega +\mu +1\right)+3\Big],\nn\\
\delta^e_n&=&\mu ^2 \left(2-16 \eta ^4 \omega ^2\right)-208 \eta ^4 \omega ^2+84 i \eta ^2 \omega +\mu  \left(-224 \eta ^4 \omega ^2+48 i \eta ^2 \omega +4\right)+\left(\mu ^2+14 \mu +22\right) n^2\nn\\
&-&2 i n \left(4 \eta ^2 \mu ^2 \omega +\mu  \left(56 \eta ^2 \omega -6 i\right)+70 \eta ^2 \omega -15 i\right)+11,\nn\\
\sigma^e_n&=&-3 \Big[-128 \eta ^4 \omega ^2+88 i \eta ^2 \omega +\mu  \left(-32 \eta ^4 \omega ^2+16 i \eta ^2 \omega +3\right)+2 (\mu +4) n^2\nn\\
&+&n \left(\mu  \left(-4-16 i \eta ^2 \omega \right)-64 i \eta ^2 \omega -22\right)+15\Big],\nn\\
\rho^e_n&=&9 \left(-4 i \eta ^2 \omega +n-2\right)^2.
\eeq
By the Gaussian elimination~\cite{Leaver:1990zz}, a three-term recurrence relation can also be obtained.
From Eq.~\eqref{alpha1} we can obtain
\be
\fc{a_1}{a_0}=-\fc{\beta_0}{\alpha_0}.\label{a1a0}
\ee
On the other hand, from the three-term recurrence relation~\eqref{alpha2} we can obtain a continued fraction
\be
\fc{a_1}{a_0}=\fc{-\delta_1}{\beta_1-\fc{\alpha_1\delta_2}{\beta_2-\fc{\alpha_2\delta_3}{\beta_3-\cdots}}}
\ee
which can be rewritten into a usual notation as
\be
\fc{a_1}{a_0}=\fc{-\delta_1}{\beta_1-}\fc{\alpha_1\delta_2}{\beta_2-}\fc{\alpha_2\delta_3}{\beta_3-}\cdots\label{a1a02}
\ee
Comparing Eqs.~\eqref{a1a0} and \eqref{a1a02}, we can obtain the QNFs by solving the following relation
\be
0=\beta_0-\fc{\alpha_0\delta_1}{\beta_1-}\fc{\alpha_1\delta_2}{\beta_2-}\fc{\alpha_2\delta_3}{\beta_3-}\cdots
\ee

\emph{Direct integration} Another useful method to solve the fundamental and lower overtone QNFs is the direct integration method~\cite{Pani:2013pma}. First we should expand the wave function $\Psi^s$ at horizon and infinity to satisfy the boundary condition~\eqref{boundary}. Then we integrate the master equation~\eqref{master equation} from both horizon and infinity and match them at a certain point. By imposing the wave functions should be continued and their first derivatives are also continued, the eigenvalues, i.e., the QNFs can be solved.

\emph{WKB method} The basic idea of the WKB method is to match the asymptotic WKB solutions at the two boundaries (horizon and infinity) with the Taylor expansion near the peak of the potential through two turning points. For 6th order WKB method, the QNFs can be evaluated through~\cite{Konoplya:2003ii}
\be
\fc{i(\omega_n^2-V_0)}{\sqrt{-2V_0''}}+\sum_{j=2}^{6}\Lambda_j=n+\fc{1}{2}, \label{wkbterm}
\ee
where $V_0$ and $V_0''$ represent the peak value of the effective potential and the second derivative with respect to the tortoise coordinate at the peak value of the effective potential, respectively. The correction terms $\Lambda_j$ depend on the peak value of the potential and higher-order derivatives of the peak value~\cite{Konoplya:2003ii,Iyer:1986np,Konoplya:2002zu}.

\begin{center}
\begin{table}[!htb]
\begin{tabular}{|c|c|c|c|c|}
\hline
$\gamma$   &~$n$~          &Direct Integration         &~Continued Fraction  &       WKB       \\
\hline
  & &$\omega_\text{R} M$~~~~$\omega_\text{I} M$~&$\omega_\text{R} M$~~~~$\omega_\text{I} M$~&$\omega_\text{R} M$~~~~~$\omega_\text{I} M$~\\
\hline
   -0.05& 0&    0.101906~~~~-0.094972&0.100185~~~~-0.095142&0.100198~~~~-0.091442\\
        & 1&    0.078359~~~~-0.368925&0.077706~~~~-0.316039&0.080754~~~~-0.312488\\
\hline
  0     & 0&    0.111714~~~~-0.104218&0.110454~~~~-0.104894&0.110464~~~~-0.100819\\
        & 1&    0.085238~~~~-0.346023&0.085671~~~~-0.348433&0.089023~~~~-0.344552\\
\hline
  0.05  & 0&    0.123085~~~~-0.115260&0.122387~~~~-0.116227&0.122401~~~~-0.111708\\
        & 1&    0.094612~~~~-0.389441&0.094927~~~~-0.386075&0.098647~~~~-0.381752\\
\hline

\end{tabular}\\
\caption{The fundamental and first overtone QNFs of the scalar field for different values of $\gamma$ through three methods. The angular number $l$ is set to $l=0$.}
\label{scalar}
\end{table}
\end{center}

\begin{center}
\begin{table}[!htb]
\begin{tabular}{|c|c|c|c|c|}
\hline
$\gamma$   &~$n$~          &Direct Integration         &~Continued Fraction  &       WKB       \\
\hline
  & &$\omega_\text{R} M$~~~~$\omega_\text{I} M$~&$\omega_\text{R} M$~~~~$\omega_\text{I} M$~&$\omega_\text{R} M$~~~~~$\omega_\text{I} M$~\\
\hline
   -0.05& 0&    0.231800~~~~-0.084056&0.231800~~~~-0.084056&0.231744~~~~-0.084175\\
        & 1&    0.199991~~~~-0.269997&0.201756~~~~-0.266227&0.201581~~~~-0.266583\\
\hline
  0     & 0&    0.248263~~~~-0.092488&0.248263~~~~-0.092488&0.248191~~~~-0.092637\\
        & 1&    0.209938~~~~-0.290036&0.214515~~~~-0.293668&0.214295~~~~-0.294118\\
\hline
  0.05  & 0&    0.266761~~~~-0.102253&0.266761~~~~-0.102252&0.266668~~~~-0.102443\\
        & 1&    0.221822~~~~-0.316663&0.228638~~~~-0.325569&0.228357~~~~-0.326147\\

\hline

\end{tabular}\\
\caption{The fundamental and first overtone QNFs of the electromagnetic field for different values of $\gamma$ through three methods. The angular number $l$ is set to $l=1$.}
\label{EM}
\end{table}
\end{center}

\begin{center}
\begin{table}[!htb]
\begin{tabular}{|c|c|c|c|c|}
\hline
$\gamma$   &~$n$~          &Direct Integration         &~Continued Fraction  &       WKB       \\
\hline
  & &$\omega_\text{R} M$~~~~$\omega_\text{I} M$~&$\omega_\text{R} M$~~~~$\omega_\text{I} M$~&$\omega_\text{R} M$~~~~~$\omega_\text{I} M$~\\
\hline
   -0.05& 0&    0.339330~~~~-0.080708&0.339329~~~~-0.080701&0.339283~~~~-0.080637\\
        & 1&    0.315003~~~~-0.249025&0.314906~~~~-0.248467&0.314536~~~~-0.248073\\
\hline
    0   & 0&    0.373672~~~~-0.088962&0.373672~~~~-0.088962&0.373619~~~~-0.088891\\
        & 1&    0.350008~~~~-0.270013&0.346711~~~~-0.273915&0.346297~~~~-0.273480\\
\hline
  0.05  & 0&    0.414630~~~~-0.098580&0.414635~~~~-0.098588&0.414579~~~~-0.098509\\
        & 1&    0.384614~~~~-0.303263&0.384808~~~~-0.303534&0.384358~~~~-0.303053\\
\hline

\end{tabular}\\
\caption{The fundamental and first overtone QNFs of the odd parity gravitational field for different values of $\gamma$ through three methods. The angular number $l$ is set to $l=2$.}
\label{gravitation}
\end{table}
\end{center}

We solve the QNFs for the perturbed fields through the above three methods. The results are shown in Table~\ref{scalar} (QNFs of the scalar field), Table~\ref{EM} (QNFs of the electromagnetic field), Table~\ref{gravitation} (QNFs of the gravitational field). The results obtained by these three methods agree well with each other, which shows that our results are correct. From the three tables we find the differences between the three methods for the fundamental QNFs are less than that for the first overtone QNFs. Especially, the fundamental QNFs of the electromagnetic field and the gravitational field solved through the direct integration method and the continued fraction method are almost the same. Besides, when the Lorentz-violating parameter $\gamma=0$, the results recover to that of the Schwarzschild black hole. When $\gamma$ deviates from zero, the QNFs will be different, but the differences are small. On the other hand, due to the modification of the KR field, the isospectrality between the even and odd parity modes is broken, which can be seen from Table~\ref{gravitation1}. The difference of the frequencies between two parity modes increases with the absolute value of $\gamma$. 

\begin{center}
\begin{table}[!htb]
\begin{tabular}{|c|c|c|c|c|c|}
\hline
   ~$\gamma$~          &-0.1         &-0.05  &       0&0.05&0.1       \\
\hline
  $\omega_\text{R} M$~~~~~odd& 0.310148&0.339329&0.373672&0.414635&0.461323\\
             ~~~~~~~~~~~ even&0.308820&0.338931&0.373672&0.414041&0.464297\\
\hline
  $\omega_\text{I} M$~~~~~odd&-0.073556&-0.080701 &-0.088962&-0.098588&-0.109904\\
     ~~~~~~~~~~~ even        &-0.073523&-0.080691&-0.088962&-0.098573&-0.109830\\
\hline
\end{tabular}\\
\caption{The fundamental QNFs of the odd parity and even parity gravitational perturbation for different values of $\gamma$. The angular number $l$ is set to $l=2$.}
\label{gravitation1}
\end{table}
\end{center}

\begin{figure*}[htb]
\begin{center}
\subfigure[~The real parts of the QNFs.]  {\label{real}
\includegraphics[width=6cm]{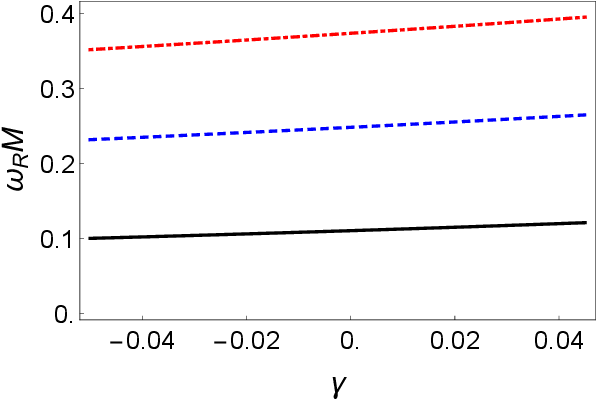}}
\subfigure[~The imaginary parts of the QNFs.]  {\label{Im}
\includegraphics[width=6cm]{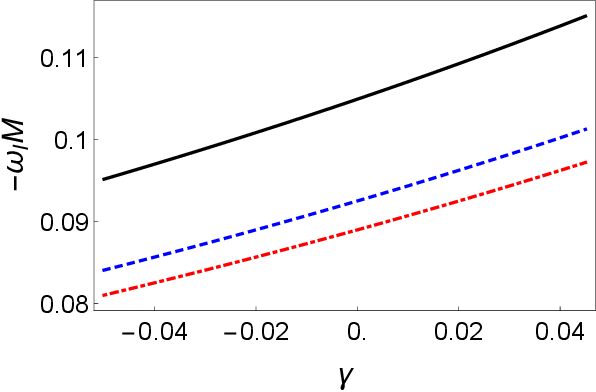}}
\end{center}
\caption{The effect of the Lorentz-violating parameter $\gamma$ on the QNFs. The black, blue dashed, and  red dot dashed lines denote the QNFs of the scalar field, electromagnetic field, and odd parity gravitational field, respectively.}
\label{realimgamma}
\end{figure*}

The effect of the Lorentz-violating parameter $\gamma$ on the QNFs is shown in Fig.~\ref{realimgamma}. From this figure we can see that the effects for the three perturbed fields are the same. Both the real parts and the absolute value of imaginary parts of the QNFs increase with parameter $\gamma$.

\begin{figure*}[htb]
\begin{center}
\subfigure[$ $]  {\label{TS}
\includegraphics[width=5.7cm]{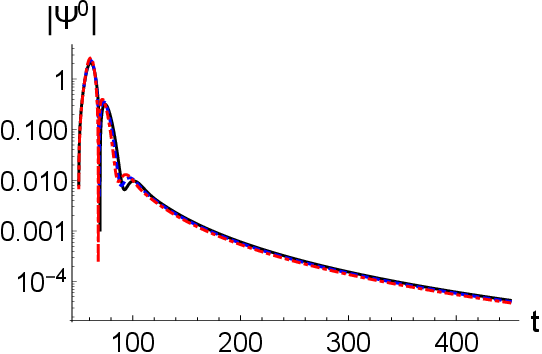}}
\subfigure[$ $]  {\label{TE}
\includegraphics[width=5.7cm]{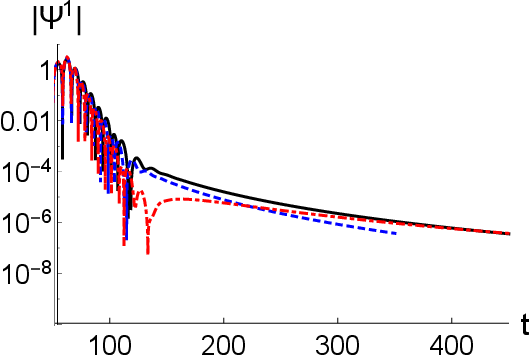}}
\subfigure[$ $]  {\label{Tg}
\includegraphics[width=5.7cm]{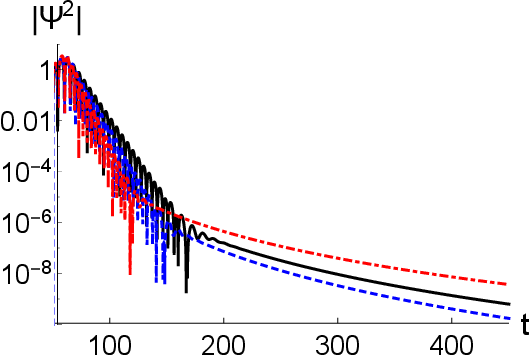}}
\end{center}
\caption{The time evolution for different perturbed fields. The mass $M$ is set to $M=1$, and the Lorentz violating parameter $\gamma$ is set to $\gamma=-0.05$ (the black lines), $\gamma=0$ (the blue dots), and $\gamma=0.05$ (the red dashed dots).  (a) The time evolution for the scalar field and $l=0$. (b) The time evolution for the Maxwell field and $l=1$. (c)  The time evolution for the odd parity gravitational field and $l=2$.}
\label{timeevolution}
\end{figure*}

The QNMs can tell us how the perturbations decay, but we do not know the amplitude of the perturbations. In order to understand the system more completely, we should investigate the time evolution of the perturbed fields. In the time domain, we will use the null coordinates $u=t-r_*$ and $v=t+r_*$ to study the time evolution. The perturbation equation in the null coordinates can be rewritten as
\be
4\fc{\partial^2\Psi^s}{\partial_u\partial_v}+V_s\Psi^s=0.
\ee
The initial data is chosen to be a Gaussian wave packet in the $v$ coordinate
\beq
\Psi^s(0,v)&=&\exp(-\fc{(v-v_c)^2}{2\sigma^2}),\nn\\
\Psi^s(u,0)&=&0.
\eeq
The Gaussian wave packet is located at $v_c=10M$, and its width is $\sigma=1M$. The ranges of the $u,v$ coordinates are both $(0,500M)$. We extract the data at $r_*=50M$. The results of the time evolution for the three perturbed fields are shown in Fig.~\ref{timeevolution}. From these figures, we can see that, the amplitude of the black lines ($\gamma=-0.05$) larger than that of the blue lines ($\gamma=0$) and the red lines ($\gamma=0.05$) in the QNMs dominate part. That is, the perturbed fields will decay more quickly as the Lorentz-violating parameter $\gamma$ increase. This is consistent with the result obtained in frequency domain. Besides, we can obtain the QNFs by fitting the data. The results agree well with that of frequency domain. For example, the fitted QNF for the black line of Fig.~\ref{Tg} (the gravitational field with $\gamma=-0.05$ and $l=2$) is $0.332431-0.081783 i$. The result obtained by the continued fraction method is $0.339329-0.080701 i$. Considering the numerical error, the two results agree well with each other.

\section{Greybody factor}\label{gbfactor}

Another very important aspect of perturbations around a black hole is the absorption cross-section. The greybody factor is defined as the probability of an outgoing wave reach to infinity or an incoming wave to be absorbed by the black hole~\cite{Konoplya:2019ppy,Cardoso:2005vb,Dey:2018cws}. Recently, it has been noted that the ringdown signal following an
extreme mass ratio merger can be modelled using the greybody factor~\cite{Oshita:2023cjz}. So, it is very important to study the greybody factor of the Lorentz-violating black hole.

We will calculate the greybody factor through the WKB method~\cite{Konoplya:2019hlu}. The boundary condition for the scattering process is different from that of the QNMs, which can be written as
\beq
\Psi^s&=&e^{-i\omega r_*}+Re^{i\omega r_*},~~~~r_*\rightarrow+\infty,\nn\\
\Psi^s&=&Te^{-i\omega r_*},~~~~~~~~~~~~~~r_*\rightarrow-\infty,\label{boungf}
\eeq
where $R$ and $T$ represent the reflection coefficient and transmission coefficient, respectively. From the boundary condition~\eqref{boungf} we can see that, there are both ingoing wave and outgoing wave at infinity, but only ingoing wave at horizon. Using the 6th order WKB method, the reflection and transmission coefficients can be obtained
\beq
|R|^2&=&\fc{1}{1+e^{-2\pi i K}},\\
|T|^2&=&\fc{1}{1+e^{2\pi i K}}=1-|R|^2,
\eeq
where $K$ is a parameter which can be obtained by the WKB formula
\be
K=\fc{i(\omega^2-V_0)}{\sqrt{-2V_0''}}-\sum_{j=2}^{6}\Lambda_j,
\ee
where the correction terms $\Lambda_j$ are the same as in Eq.~\eqref{wkbterm} and are all imaginary numbers.

\begin{figure*}[htb]
\begin{center}
\subfigure[~The greybody factor for the scalar field.]  {\label{gbscalar}
\includegraphics[width=5.5cm]{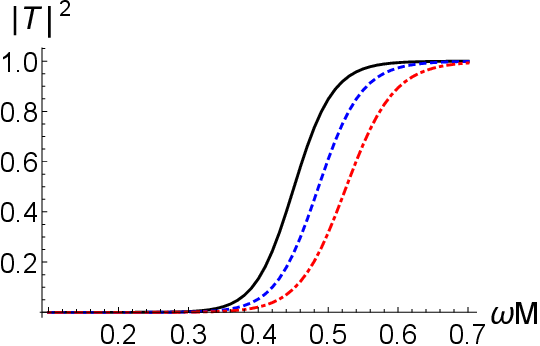}}
\subfigure[~The greybody factor for the electromagnetic field.]  {\label{gbEM}
\includegraphics[width=5.5cm]{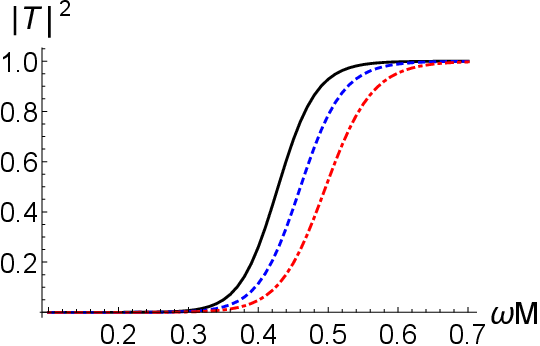}}
\subfigure[~The greybody factor for the odd parity gravitational field.]  {\label{gbgr}
\includegraphics[width=5.5cm]{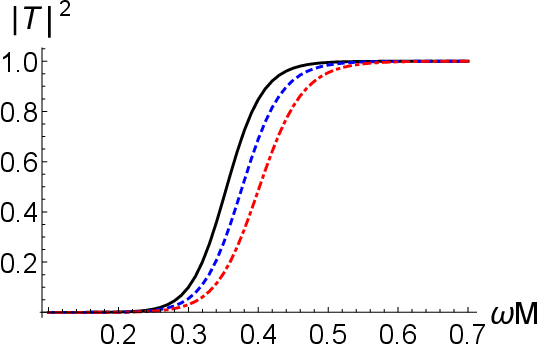}}
\end{center}
\caption{The effect of the Lorentz-violating parameter $\gamma$ on the greybody factor for three perturbed fields. The black lines denote the greybody factor with $\gamma=-0.05$, the blue dashed lines denote the greybody factor with $\gamma=0$, and the red dot dashed lines denote the greybody factor with $\gamma=0.05$.}
\label{greybody}
\end{figure*}

The greybody factor for the three perturbed fields are plotted in Fig.~\ref{greybody}. From this figure we can see that, the greybody factor decrease with the parameter $\gamma$, which means that with the increasing of $\gamma$, a smaller fraction of the perturbed field can penetrate the potential barrier. This is consistent with the effective potential, where the height of the potential barrier increase with the parameter $\gamma$, which can be seen from Fig.~\ref{plotpotential}.

\section{Conclusions}\label{conclusion}

In this paper we studied the QNMs and the greybody factor in the background of a spherically symmetric Lorentz-violating black hole. The Lorentz symmetry is spontaneously broken due to the vacuum condensate of the KR field. The Lorentz-violating effect is represented by a parameter $\gamma$, which is named as the Lorentz-violating parameter.

We investigated three kinds of perturbed fields, the scalar field, electromagnetic field, and gravitational field. The perturbation equations for the perturbations can be written into a Schr\"{o}dinger equation~\eqref{master equation}. The effective potentials were plotted in Fig.~\ref{plotpotential}, from which we can see that the height of the effective potential barrier increases with the Lorentz-violating parameter $\gamma$. After choosing the boundary condition, the QNMs become a eigenvalue problem. We used three methods to solve the QNFs, i.e. the continued fraction method, direct integration method, and WKB method. The results are shown in Tables~\ref{scalar}, \ref{EM}, and~\ref{gravitation}. The QNFs obtained from the three methods agree well with each other, which confirms our results are correct. The effect of the Lorentz-violating parameter $\gamma$ on the QNFs is shown in Fig.~\ref{realimgamma}. From this figure we showed that both the real parts and the absolute value of the imaginary parts of the QNFs increase with $\gamma$. Besides, the isospectrality between the odd and even modes of the gravitational perturbation is broken, and the difference of the frequencies between two parity modes increases with the absolute value of $\gamma$. For completeness, we studied the time evolution of the perturbed fields starting with a Gaussian wave packet. The results are shown in Fig.~\ref{timeevolution}. From the numerical data, we fitted the QNFs, the results agree well with that of the frequency domain. Figure~\ref{timeevolution} shows that the perturbed fields will decay more quickly as the Lorentz-violating parameter $\gamma$ increase.

Using the 6th order WKB method, we also calculated the greybody factor for these three perturbed fields. The results are shown in Fig.~\ref{greybody}. The effect of the Lorentz-violating parameter $\gamma$ shows that, the greybody factor decrease with the parameter $\gamma$. This is consistent with the effective potential.

It is known that using the spectroscopy analysis could estimate the physical parameters of the black hole~\cite{Berti:2005ys,Berti:2016lat,Berti:2018vdi,Zhao:2022lrl,Dreyer:2003bv}, and the overtone modes are very important for determining the remnant black hole parameters~\cite{Giesler:2019uxc}. However, the signal-to-noise ratio of the current ground-based gravitational wave detectors is not high~\cite{Cotesta:2022pci}, we expect that the future space-based gravitational detectors~\cite{Ruan:2018tsw,Lu:2019log,Moore:2014lga,Gong:2021gvw,Zhang:2021kkh} can help us to observe the effect of the Lorentz-violating parameter $\gamma$.

\section{Acknowledgments}

We thank Ke Yang for very useful discussion on the Lorentz-violating black hole. This work was supported by National Key Research and Development Program of China (Grant No. 2021YFC2203003), the National Natural Science Foundation of China (Grants No. 12205129, No. 12147166, No. 11875151, and No. 12247101, No. 12347111), the China Postdoctoral Science Foundation (Grant No. 2021M701529 and No. 2023M741148), the 111 Project (Grant No. B20063), the Department of education of Gansu Province: Outstanding Graduate ``Innovation Star'' Project (Grant No. 2023CXZX-057), the Major Science and Technology Projects of Gansu Province, and Lanzhou City's scientific research funding subsidy to Lanzhou University.

%

%
%

\end{document}